\documentclass[11pt,letterpaper]{article}
\usepackage[margin=1in]{geometry}
\usepackage{amsmath,amssymb,amsthm}
\usepackage{graphicx}
\usepackage{booktabs,tabularx}
\usepackage{xcolor}
\usepackage{enumitem}
\usepackage[hidelinks]{hyperref}
\newtheorem{theorem}{Theorem}[section]
\newtheorem{lemma}[theorem]{Lemma}
\newtheorem{proposition}[theorem]{Proposition}
\newtheorem{corollary}[theorem]{Corollary}
\theoremstyle{definition}
\newtheorem{definition}[theorem]{Definition}
\newtheorem{assumption}{Assumption}
\theoremstyle{remark}
\newtheorem{remark}[theorem]{Remark}

\newcommand{\sgn}{\operatorname{sgn}}
\newcommand{\E}{\mathbb{E}}
\newcommand{\Prob}{\mathbb{P}}
\newcommand{\KL}{\mathrm{KL}}
\newcommand{\Var}{\operatorname{Var}}
\newcommand{\ctx}{c_{\mathrm{tx}}}

\title{Learning AI Agents, Market Makers, and Stress in Automated Markets}
\author{Irene Aldridge\thanks{Contact: \texttt{irene.aldridge@gmail.com}.}\\[2pt]
\normalsize RiskAICenter and Cornell Tech, Cambridge and Johns Hopkins}
\date{}

\begin{document}
\maketitle
\thispagestyle{empty}

\begin{abstract}
Automated markets increasingly route order flow from learning trading agents through automated market makers (AMMs). We model trader archetypes as AI agents defined by their reward functions and analyze
the feedback they create on a constant-product AMM. Herd-dominated and arbitrage-dominated markets of equal size have identical return distributions and turnover, so no rule based on volatility,
value-at-risk, stress frequency, or volume can tell them apart; yet their stress episodes differ in length by a factor unbounded in the feedback-to-noise ratio $\beta/s$. Serial statistics identify the
feedback sign from $\Theta((s/\beta)^2\log(1/\delta))$ observations, and this rate is tight. When agents trade together, profit-seeking agents ride herds rather than lean against them, leading  to all-or-nothing participation. Spread-earning market makers restore stability but make liquidity fragile, with discontinuous collapse and hysteresis; maker--taker pricing, not a uniform fee, is the effective instrument. The results explain patterns in companion empirical work on learned
agents in cryptocurrency markets.
\end{abstract}


\newpage
\setcounter{page}{1}

\section{Introduction}\label{sec:intro}

A growing share of trading is done by automated, increasingly autonomous agents, and a growing share of matching, above all in digital assets, happens against automated market makers: deterministic pricing formulas such as the constant-product rule~\cite{uniswapv2,angeris2020}. Arbitrageurs, retail trend-followers, and front-runners~\cite{daian2020} all trade in the same pools, and each leaves a different footprint on the frequency and persistence of stress. As AI agents take on these roles, proportionate regulation requires knowing which objectives stabilize a market and whether a regulator can determine this from data.

Two obstacles stand in the way. Behavior types are not labeled in public data, and the statistics regulators usually monitor (volatility, value-at-risk, the frequency of large losses, turnover) are
properties of the \emph{one-period} return distribution. We show that such statistics are provably
blind to the property that matters most, identify what is not blind to it at an optimal rate, and
analyze what learning agents and market makers do when they trade together. We take an objective-first
view, following the empirical framework of~\cite{aldridge2026agentic}: each agent is defined by the
reward it maximizes, so assumptions are explicit while strategy is emergent.

\subsection{Our results}\label{sec:results}

We model the pool's log return as $r_t=I_t+\eta_t$, where $I_t$ is the price impact of the period's net
order flow and $\eta_t$ is exogenous news with scale $s$. Herd-followers trade in the direction of the
last move, which we show is optimal under the herd-follower's reward (Lemma~\ref{lem:herd-policy}). In
the simplest case $r_t=\beta\,\sgn(r_{t-1})+\eta_t$, where $\beta>0$ is net positive feedback
(herding) and $\beta<0$ net negative feedback (arbitrage).

\paragraph{Volatility cannot see feedback (Section~\ref{sec:blind}).} Markets with feedback $+\beta$ and
$-\beta$ have the same one-period return law and the same turnover, but their expected stress-episode
lengths differ by $G\big(\tfrac{c+\beta}{s}\big)/G\big(\tfrac{c-\beta}{s}\big)$, which is unbounded in $\beta/s$ (Theorem~\ref{thm:main}). Therefore, every functional of the return marginal and turnover treats a destabilizing and a stabilizing market identically (Corollary~\ref{cor:impossible}). For Gaussian news with $\beta=s$, episodes last $6.3$ periods under herding and $1.2$ under arbitrage
(Figure~\ref{fig:theory}a).

\paragraph{Serial statistics see it at an optimal rate.} Autocorrelation and the variance ratio have
the sign of $\beta$ (Proposition~\ref{prop:serial}). A sign-agreement test identifies $\sgn(\beta)$ with
error $\delta$ from $O((s/\beta)^2\log(1/\delta))$ observations, and no test using the full return path
needs fewer than $\Omega((s/\beta)^2\log(1/\delta))$ (Theorem~\ref{thm:detect}). A herd-follower's own
stress episodes, by contrast, last at most two periods on every path (Proposition~\ref{prop:herd-own}),
so strategy-level measurements see herders as brief sources of stress and miss their effect on market
persistence.

\paragraph{Agents playing together (Section~\ref{sec:together}).} On a path-independent pool, a
trader's own impact cancels from its round-trip payoff (Lemma~\ref{lem:payoff}); only trading ahead of
predictable herd flow pays. Leaning against the herd is therefore strictly dominated, riding with it
is a strategic complement, and under any sign-preserving learning dynamics, including replicator
dynamics and softmax policy gradient, participation tips to zero or to full depending on the starting
point; full participation strictly lengthens stress, and no uniform fee can induce stabilizing behavior
(Theorem~\ref{thm:together}, Corollary~\ref{cor:fee}).

\paragraph{Market makers (Section~\ref{sec:mm}).} Spread-earning market makers enter the price process
exactly as negative riders, so persistence depends only on the net index $z=w-\sigma h$ of rider capital
minus absorbed flow. They supply the negative feedback fee-paying traders will not, but liquidity
becomes fragile: when rider capital crosses a threshold $\bar X_c$ market makers exit at once and
persistence jumps, and they return only below $\bar X_c-\bar\sigma h$, so the hysteresis band is exactly
as wide as their absorption capacity (Theorem~\ref{thm:mm}, Figure~\ref{fig:theory}b). A maker rebate
widens the stable region; a taker fee removes riders (Corollary~\ref{cor:makertaker}).

\paragraph{Techniques.} Three observations drive the analysis. (i) With symmetric noise, the return
sign is a Markov chain whose flips are i.i.d.\ and independent of the current sign
(Lemma~\ref{lem:chain}), which decouples the marginal (a function of $|\beta|$) from the serial
structure (a function of $\sgn(\beta)$). (ii) On a path-independent pool, the own-impact part of a
round trip telescopes away in expectation (Lemma~\ref{lem:payoff}), so payoffs depend on aggregate
behavior only through one persistence probability $q$. (iii) $q$ is monotone in each population's net
flow, which makes the interaction game one of strategic complements within groups and substitutes
across groups~\cite{milgrom1990}, so equilibria, stability, and learning basins follow from monotone
comparative statics.

\paragraph{Empirical connection.} Section~\ref{sec:empirical} relates the results to the companion
study~\cite{aldridge2026agentic}, which trains these archetypes with deep reinforcement learning on
hourly cryptocurrency data. The theory explains its main strategy-level patterns and shows which of
them do not survive interaction.

\subsection{Related work}\label{sec:related}

De~Long, Shleifer, Summers, and Waldmann~\cite{dssw1990} show that rational speculators may amplify
positive-feedback traders; Theorem~\ref{thm:together} derives this for learning agents on an AMM, with
equilibrium tipping and learning basins. Agent-based finance shows that heterogeneous traders produce
fat tails and crashes~\cite{lebaron2006}. Classical microstructure explains how liquidity providers set
prices~\cite{kyle1985,glosten1985} and absorb temporary imbalances~\cite{grossman1988}; liquidity can
evaporate in spirals~\cite{brunnermeier2009}, as in the 2010 Flash Crash~\cite{kirilenko2017}. Our
market-maker result gives a closed-form collapse threshold and hysteresis band. Constant-function
market makers are analyzed in~\cite{angeris2020}, arbitrage costs against AMMs in~\cite{milionis2022},
and front-running on decentralized exchanges in~\cite{daian2020}; our execution model follows frequent
batch auctions~\cite{bcs2015}. Learning algorithms can reach unintended outcomes such as tacit
collusion~\cite{calvano2020}; our learning results use sign-preserving population
dynamics~\cite{sandholm2010} and games with strategic complementarities~\cite{milgrom1990}. The
variance-ratio test of~\cite{lo1988} is a serial statistic of the kind Theorem~\ref{thm:detect} shows a
regulator needs.

\paragraph{Organization.} Section~\ref{sec:archetypes} recalls the archetypes' rewards and
Section~\ref{sec:model} gives the AMM feedback model. Section~\ref{sec:blind} proves the impossibility
and detection results, and Sections~\ref{sec:together} and~\ref{sec:mm} analyze agents trading
together and market makers. Section~\ref{sec:empirical} connects the results to data,
Section~\ref{sec:open} discusses implications for regulators, and Appendix~\ref{app:numerics} verifies every closed form by simulation.

\section{Trader archetypes as reward functions}\label{sec:archetypes}

We recall the archetypes of~\cite{aldridge2026agentic}. Trading is a discrete exposure-control problem. At each step $t$, the agent observes a feature vector $o_t$ and chooses an action in $\{\text{sell},\text{hold},\text{buy}\}$, which moves its exposure
$e_t\in\{0,\tfrac12,1\}$ by $-\Delta$, $0$, or $+\Delta$ with $\Delta=\tfrac12$. Exposure $0$ is flat,
$\tfrac12$ half invested, and $1$ fully long; there is no short selling. The base reward is a
volatility-normalized net return with a stress penalty:
\begin{equation}\label{eq:base-reward}
r^{\mathrm{base}}_t=\frac{e_tR_{t+1}-\ctx\,|e_t-e_{t-1}|}{\hat\sigma_t+\epsilon}
-\lambda_{\mathrm{stress}}\,\mathbf 1[\mathrm{stress}_t]\,e_t ,
\end{equation}
where $R_{t+1}$ is the next-period asset return, $\ctx$ the proportional transaction cost,
$\hat\sigma_t$ trailing volatility, $\epsilon>0$ a small constant, and $\mathrm{stress}_t$ a market
crash-regime indicator. Each archetype modifies~\eqref{eq:base-reward}:

\begin{itemize}[leftmargin=1.2em]
\item \textbf{Rational arbitrageur.} Maximizes risk-adjusted return with additional penalties on
realized volatility and turnover. It acts conservatively and reduces exposure as risk rises.
\item \textbf{Manipulator.} A stylized proxy for front-running and wash trading. It receives a bonus
$\kappa_{\mathrm{turn}}|e_t-e_{t-1}|$ and a short discount horizon, which favors rapid position cycling.
\item \textbf{Retail herd-follower.} Mimics trend-chasing and fear of missing out. It receives a bonus
for aligning exposure with recent momentum,
\begin{equation}\label{eq:herd-bonus}
\kappa_{\mathrm{mom}}\,\sgn(\mathrm{mom}_t)\,(e_t-\tfrac12).
\end{equation}
\end{itemize}

Sections~\ref{sec:model}--\ref{sec:mm} ask what these objectives do to a market when the agents'
trades move prices; the companion study~\cite{aldridge2026agentic} trains them on market data.

\section{A feedback model of an automated market}\label{sec:model}

\paragraph{The AMM.} A constant-product pool with reserves $(x,y)$ quotes price $p=y/x$. Depositing
$d$ units of $y$ moves the reserves to $(x',y+d)$ with $x'(y+d)=xy$, so
\[
\log p'-\log p=2\log\!\Big(1+\frac{d}{y}\Big)=\frac{2d}{y}+O\!\Big(\frac{d^2}{y^2}\Big).
\]
Because the pool is path-independent, a period's orders aggregate into one net flow $N_t$, measured in
units of the $y$-reserve. We use the first-order impact $\lambda N_t$ with $\lambda=2$;
Remark~\ref{rem:exact} treats the exact impact.

\paragraph{Order flow.} Time is discrete, $t=1,2,\dots$, and $r_t=\log p_t-\log p_{t-1}$ is the pool's return. Each period brings exogenous news $\eta_t$ (for example, moves in an external reference market
transmitted to the pool) and endogenous flow from two populations:
\begin{itemize}[leftmargin=1.2em]
\item mass $h\ge0$ of \emph{herd-followers}, each buying one unit if the last return was positive and selling one unit otherwise; and
\item mass $\alpha\ge0$ of \emph{negative-feedback traders}, each trading one unit against the last return.
\end{itemize}
Writing $\beta=\lambda(h-\alpha)$ and $D_t=\sgn(r_t)$, the pool's return is
\begin{equation}\label{eq:model}
r_t=\beta D_{t-1}+\eta_t .
\end{equation}

\begin{assumption}\label{ass:noise}
The $\eta_t$ are i.i.d.\ and independent of $r_0$, with $\eta_t=s\,\xi_t$ for $s>0$, where $\xi_t$ has a continuous, strictly increasing distribution function $G$ that is symmetric about zero
($G(-x)=1-G(x)$). The initial sign $D_0$ is uniform on $\{\pm1\}$.
\end{assumption}

\begin{definition}[Stress]\label{def:stress}
For a threshold $c\ge0$, period $t$ is \emph{stressed} if $S_t=\mathbf 1[r_t<-c]$ equals one. A \emph{stress episode} is a maximal run of consecutive stressed periods, and $L$ denotes its length. The
\emph{stress frequency} is $F=\Prob(S_t=1)$, the \emph{volatility} is $\sqrt{\Var(r_t)}$, and the \emph{gross turnover} is $h+\alpha$ units per period.
\end{definition}

\paragraph{From rewards to feedback.} The next lemma shows that the herd-follower's reward induces the herd rule discussed above.

\begin{lemma}[The herd reward induces sign-momentum]\label{lem:herd-policy}
Let $\mu_t=\E[R_{t+1}\mid\mathcal F_t]$, and suppose the agent is myopic (discount factor $0$) with $e_{t-1}\in(0,1)$. If
\[
\kappa_{\mathrm{mom}}>\frac{|\mu_t|+\ctx}{\hat\sigma_t+\epsilon}+\lambda_{\mathrm{stress}},
\]
then the unique expected-reward-maximizing action under~\eqref{eq:base-reward}--\eqref{eq:herd-bonus}
is \emph{buy} when $\mathrm{mom}_t>0$ and \emph{sell} when $\mathrm{mom}_t<0$.
\end{lemma}

\begin{proof}
Write the action as $j\in\{-1,0,1\}$, so $e=e_{t-1}+j/2$, and let $\chi_t=\sgn(\mathrm{mom}_t)$. The
expected one-step reward is $eK-\frac{\ctx|j|}{2(\hat\sigma_t+\epsilon)}$ plus terms that do not depend
on $j$, where
$K=\frac{\mu_t}{\hat\sigma_t+\epsilon}-\lambda_{\mathrm{stress}}\mathbf 1[\mathrm{stress}_t]+\kappa_{\mathrm{mom}}\chi_t$.
Relative to holding, buying gains $\tfrac12\big(K-\tfrac{\ctx}{\hat\sigma_t+\epsilon}\big)$ and selling
gains $\tfrac12\big(-K-\tfrac{\ctx}{\hat\sigma_t+\epsilon}\big)$. If $\chi_t=1$, the hypothesis gives
$K\ge\kappa_{\mathrm{mom}}-\frac{|\mu_t|}{\hat\sigma_t+\epsilon}-\lambda_{\mathrm{stress}}>\frac{\ctx}{\hat\sigma_t+\epsilon}>0$.
Therefore, buying beats holding, and buying also beats selling by $K>0$. If $\chi_t=-1$, then $-K\ge\kappa_{\mathrm{mom}}-\frac{|\mu_t|}{\hat\sigma_t+\epsilon}>\frac{\ctx}{\hat\sigma_t+\epsilon}$,
so selling is the unique maximizer.
\end{proof}

With a one-period momentum signal, a population of such agents emits a net flow $h\,D_{t-1}$, the herd term in~\eqref{eq:model}. In Section~\ref{sec:blind}, the negative-feedback mass $\alpha$ is a modeling device; Section~\ref{sec:together} asks whether any profit-seeking agent would choose to lean, and Section~\ref{sec:mm} shows who does.

\section{Volatility cannot see feedback}\label{sec:blind}

\begin{lemma}[The sign process]\label{lem:chain}
Let $q=G(\beta/s)$. Under Assumption~\ref{ass:noise}, the flips $A_t=D_tD_{t-1}$, $t\ge1$, are i.i.d.\ with $\Prob(A_t=1)=q$ and independent of $D_0$. Consequently, $(D_t)$ is a stationary symmetric Markov chain with uniform marginals.
\end{lemma}

\begin{proof}
$A_t=\sgn(D_{t-1}r_t)=\sgn(\beta+D_{t-1}\eta_t)$. Conditional on $\mathcal F_{t-1}$, the variable $D_{t-1}\eta_t$ has the law of $\eta_t$, by independence and symmetry. So $A_t$ is independent of
$\mathcal F_{t-1}$ with $\Prob(A_t=1)=\Prob(\eta_t>-\beta)=G(\beta/s)$. Since $D_t=D_0\prod_{k\le t}A_k$ and $D_0$ is uniform, each $D_t$ is uniform.
\end{proof}

\begin{theorem}[Volatility equivalence, persistence separation]\label{thm:main}
Under Assumption~\ref{ass:noise}, for every $t\ge1$ and every $c\ge0$:
\begin{enumerate}[label=(\roman*)]
\item The law of $r_t$ is $\tfrac12G_s(\cdot-\beta)+\tfrac12G_s(\cdot+\beta)$, where $G_s(x)=G(x/s)$.
It depends on $\beta$ only through $|\beta|$. In particular,
$\Var(r_t)=\beta^2+s^2\Var(\xi_1)$ when $\xi_1$ has finite variance, and
\[
F(\beta)=\tfrac12\Big[G\big(\tfrac{-c-\beta}{s}\big)+G\big(\tfrac{\beta-c}{s}\big)\Big]=F(-\beta).
\]
\item Stress-episode lengths are i.i.d.\ geometric with $\Prob(L>k)=G\big(\tfrac{\beta-c}{s}\big)^k$,
so
\[
\E[L](\beta)=\frac{1}{G\big(\tfrac{c-\beta}{s}\big)},
\]
which is strictly increasing in $\beta$.
\item For $\beta>0$,
\[
\frac{\E[L](\beta)}{\E[L](-\beta)}=\frac{G\big(\tfrac{c+\beta}{s}\big)}{G\big(\tfrac{c-\beta}{s}\big)}>1,
\]
and the ratio tends to $\infty$ as $\beta/s\to\infty$.
\end{enumerate}
\end{theorem}

\begin{proof}
(i) By Lemma~\ref{lem:chain}, $D_{t-1}$ is uniform and independent of $\eta_t$, so
$r_t=\beta D_{t-1}+\eta_t$ is an equal mixture of $\eta_t+\beta$ and $\eta_t-\beta$. Replacing $\beta$
by $-\beta$ swaps the components. The variance follows because $D_{t-1}$ and $\eta_t$ are independent
with mean $0$, and $F$ is the mixture's distribution function at $-c$.

(ii) If $S_t=1$, then $r_t<-c\le0$, so $D_t=-1$. Hence
$\Prob(S_{t+1}=1\mid\mathcal F_t,S_t=1)=\Prob(-\beta+\eta_{t+1}<-c)=G\big(\tfrac{\beta-c}{s}\big)$,
whatever the history. Each period of an episode continues it independently with this probability, so
$L$ is geometric with $\E[L]=1/\big(1-G(\tfrac{\beta-c}{s})\big)=1/G(\tfrac{c-\beta}{s})$, which is
strictly increasing in $\beta$ because $G$ is.

(iii) Apply (ii) at $\pm\beta$. As $\beta/s\to\infty$, the numerator tends to $1$ and the denominator
to $0$.
\end{proof}

For Gaussian news and $c=0$, $\E[L](\beta)=1/\Phi(-\beta/s)$: $\beta=s$ gives expected episodes of
$6.3$ periods, against $1.19$ for $\beta=-s$, with identical volatility and stress frequency
(Figure~\ref{fig:theory}a).

\begin{corollary}[Marginal-based rules cannot separate feedback]\label{cor:impossible}
Let $\Psi$ be any functional of the one-period return law and the gross turnover: volatility,
value-at-risk, expected shortfall, stress frequency at any threshold, kurtosis, or a volume cap. Then
$\Psi$ takes the same value on the herd-dominated market $(h,\alpha)=(m,0)$ and the
arbitrage-dominated market $(h,\alpha)=(0,m)$, yet their expected stress-episode lengths differ by the
factor in Theorem~\ref{thm:main}(iii), which is unbounded in $\lambda m/s$.
\end{corollary}

\begin{proof}
The two markets have $\beta=\pm\lambda m$ and the same gross turnover $m$. Apply
Theorem~\ref{thm:main}(i) and (iii).
\end{proof}

A regulator must therefore look at serial structure. The simplest serial statistic already works.

\begin{proposition}[Serial statistics see feedback]\label{prop:serial}
If, in addition, $\xi_1$ has finite variance, then $\E[r_t]=0$ and
$\mathrm{Cov}(r_t,r_{t-1})=\beta\,\E|r_{t-1}|$. Hence the first-order autocorrelation $\rho_1$ and the
two-period variance ratio $\mathrm{VR}(2)-1=\rho_1$~\cite{lo1988} have the sign of $\beta$.
\end{proposition}

\begin{proof}
$r_t$ is a symmetric mixture by Theorem~\ref{thm:main}(i), so $\E[r_t]=0$. Since $\eta_t$ is
independent of $r_{t-1}$ with mean $0$, $\E[r_tr_{t-1}]=\beta\,\E[D_{t-1}r_{t-1}]=\beta\,\E|r_{t-1}|$,
and $\E|r_{t-1}|>0$ because $r_{t-1}$ has a continuous law. For a stationary series,
$\mathrm{VR}(2)=\Var(r_t+r_{t-1})/(2\Var(r_t))=1+\rho_1$.
\end{proof}

The next theorem gives the exact cost of reading the sign.

\begin{theorem}[Tight sample complexity of detecting feedback]\label{thm:detect}
Assume Gaussian news ($G=\Phi$) and $0<|\beta|\le s$. Given $r_0,\dots,r_n$, the test that outputs
``herding'' iff $\hat q=\frac1n\sum_{t=1}^n\mathbf 1[D_t=D_{t-1}]>\tfrac12$ identifies $\sgn(\beta)$
with error at most $\delta$ whenever
\[
n\ge9\Big(\frac{s}{\beta}\Big)^2\log\frac1\delta .
\]
Conversely, for $\delta<1/4$, every test that observes the full returns has error larger than $\delta$
under $\beta$ or under $-\beta$ unless
\[
n\ge\frac12\Big(\frac{s}{\beta}\Big)^2\log\frac1{4\delta}.
\]
\end{theorem}

\begin{proof}
\emph{Upper bound.} By Lemma~\ref{lem:chain}, $n\hat q\sim\mathrm{Bin}(n,q)$ with $q=\Phi(\beta/s)$. Say
$\beta>0$; the other case is symmetric. Hoeffding's inequality~\cite{hoeffding1963} gives
$\Prob(\hat q\le\tfrac12)\le\exp(-2n(q-\tfrac12)^2)$. For $0<x\le1$,
$\Phi(x)-\tfrac12=\int_0^x\varphi\ge x\varphi(1)$, and $2\varphi(1)^2>1/9$. Hence the error is at most
$\exp(-n\beta^2/(9s^2))\le\delta$.

\emph{Lower bound.} Let $P_\pm$ be the laws of $(r_0,\dots,r_n)$ under $\pm\beta$. The law of $r_0$ is
the same under both (Theorem~\ref{thm:main}(i)). Conditional on the past, $r_t\sim N(\pm\beta
D_{t-1},s^2)$: two normals whose means differ by $2|\beta|$, so each conditional KL divergence equals
$2\beta^2/s^2$, and by the chain rule $\KL(P_+\|P_-)=2n\beta^2/s^2$. The Bretagnolle--Huber
inequality~\cite{tsybakov2009} gives, for any test $\psi$,
$P_+(\psi\ne+)+P_-(\psi\ne-)\ge\tfrac12e^{-2n\beta^2/s^2}$. So the larger of the two errors is at least
$\tfrac14e^{-2n\beta^2/s^2}$, which exceeds $\delta$ unless $n\ge\frac{s^2}{2\beta^2}\log\frac1{4\delta}$.
\end{proof}

The sign-agreement test reads only $D_t$, yet it matches the full-information lower bound up to a
constant. This is the theoretical basis for using learned policies as measurement devices: a policy
whose action depends on $\sgn(\mathrm{mom}_t)$ is, in effect, this test.

\begin{proposition}[A herd-follower's own stress is short]\label{prop:herd-own}
Consider a herd-follower with one-period momentum, exposure $e_t\in\{0,\Delta,\dots,1\}$, step $\Delta$,
net return $g_{t+1}=e_tR_{t+1}-\ctx|e_t-e_{t-1}|$, and strategy stress $\mathbf 1[g_{t+1}<-\theta]$ with
$\theta>\ctx\Delta$. On \emph{every} return path, its stress episodes last at most $1/\Delta$ periods.
With $\Delta=\tfrac12$, that is two periods.
\end{proposition}

\begin{proof}
Strategy stress at $t+1$ requires $e_tR_{t+1}<-\theta+\ctx\Delta<0$, so $e_t>0$ and $R_{t+1}<0$, and the
agent then sells: $e_{t+1}=e_t-\Delta$. Exposure therefore falls by $\Delta$ in each stressed period,
and stress is impossible at zero exposure, where $g\ge-\ctx\Delta>-\theta$.
\end{proof}

Proposition~\ref{prop:herd-own} and Theorem~\ref{thm:main}(ii) together explain a pattern found
empirically in~\cite{aldridge2026agentic} (Section~\ref{sec:empirical}). At the strategy level, herd-followers show frequent but brief stress. At the
market level, more herding \emph{lengthens} stress episodes, because $\E[L]$ increases in $h$. Strategy-level
measurements capture the former; regulation should care about the latter.

\begin{remark}[Exact constant-product impact]\label{rem:exact}
Let each population trade a net $m<1$ units of the $y$-reserve per period. With the exact impact
$2\log(1+N)$, a buy shifts the return by $\beta_+=2\log(1+m)$ and a sell by $-\beta_-$, where
$\beta_-=-2\log(1-m)>\beta_+$. The sign chain stays Markov, and the proof of
Theorem~\ref{thm:main}(ii) goes through state by state. In the herd market the continuation
probability of a stress episode is $G\big(\tfrac{\beta_--c}{s}\big)$; in the arbitrage market it is
$G\big(\tfrac{-\beta_+-c}{s}\big)$. The ratio in part (iii) becomes
$G\big(\tfrac{c+\beta_+}{s}\big)/G\big(\tfrac{c-\beta_-}{s}\big)$, which is larger still. Part (i) and
Corollary~\ref{cor:impossible} hold up to $O(m^2)$ terms.
\end{remark}

\section{Agents playing together}\label{sec:together}

Section~\ref{sec:blind} fixed the mass of each feedback type. We now let profit-seeking agents choose
their behavior in a pool where herd-followers already trade, and ask what learning converges to.

\paragraph{Players.} A mass $h>0$ of herd-followers trades $h\,D_{t-1}$ every period; by
Lemma~\ref{lem:herd-policy}, a large enough momentum bonus makes this their behavior whatever it earns.
A continuum of nonatomic \emph{rational} traders with total mass $\bar X$ each choose one of three
rules: \emph{ride} ($M$), opening a one-unit position in direction $D_{t-1}$; \emph{lean} ($C$),
opening a one-unit position in direction $-D_{t-1}$; or \emph{stay out} ($0$). Positions opened in
batch $t$ are closed in batch $t+1$, and each round trip pays a fee $\kappa>0$. Write $x_M,x_C$ for the
masses choosing $M$ and $C$, and $w=x_M-x_C$.

\paragraph{Prices.} The batch-$t$ net flow is the herd flow plus the rational traders' openings minus
their closings, with impact
\[
I_t=\lambda hD_{t-1}+\lambda w\,(D_{t-1}-D_{t-2}),\qquad r_t=I_t+\eta_t .
\]
Orders within a batch execute in uniformly random order against linear impact, so each order's
expected fill is the batch midpoint $\bar P_t=\log p_{t-1}+I_t/2$, in the spirit of batch
auctions~\cite{bcs2015}. A rider earns $D_{t-1}(\bar P_{t+1}-\bar P_t)$, and a leaner earns its
negative.

\paragraph{The sign process.} Call period $t$ \emph{continuing} if $D_{t-1}=D_{t-2}$ and \emph{fresh}
otherwise. The sign persists with probability
\[
a=G\big(\tfrac{\lambda h}{s}\big)\ \text{in continuing periods},\qquad
b=G\big(\tfrac{\lambda(h+2w)}{s}\big)\ \text{in fresh periods},
\]
because the rational openings and closings add $2\lambda wD_{t-1}$ only right after a flip. Whether a
period is fresh is therefore a two-state Markov chain, and the stationary probability that the sign
persists is
\begin{equation}\label{eq:q}
q(h,w)=\frac{b}{1-a+b},
\end{equation}
which is strictly increasing in $w$ and reduces to $q=a$, the stay probability of
Lemma~\ref{lem:chain}, when $w=0$.

\begin{lemma}[Own impact cancels]\label{lem:payoff}
In the stationary regime, the expected per-period payoffs are
\[
u_M=\lambda h\,q(h,w)-\kappa,\qquad u_C=-\lambda h\,q(h,w)-\kappa,\qquad u_0=0 .
\]
They depend on the rational traders' own flow only through $q$.
\end{lemma}

\begin{proof}
$\bar P_{t+1}-\bar P_t=I_t/2+\eta_t+I_{t+1}/2$. Let $\rho=\E[D_tD_{t-1}]=2q-1$; by stationarity it is
also $\E[D_{t-1}D_{t-2}]$. Then
\[
\E[D_{t-1}I_t]=\lambda h+\lambda w(1-\rho),\qquad\E[D_{t-1}I_{t+1}]=\lambda h\rho+\lambda w(\rho-1),
\]
and $\E[D_{t-1}\eta_t]=0$ because $\eta_t$ is independent of $\mathcal F_{t-1}$. Averaging the two, the
$w$ terms cancel, and the expected gross payoff from riding is $\lambda h(1+\rho)/2=\lambda hq$. Leaning
is the opposite position. Subtracting the fee gives both formulas.
\end{proof}

The $w$ terms cancel because a round trip on a path-independent pool cannot profit from its own
impact. A rational trader profits only by trading ahead of predictable herd flow.

\begin{theorem}[Rational agents join the herd; learning tips]\label{thm:together}
Let $u_M(x)=\lambda h\,q(h,x)-\kappa$ with $x_C=0$, and suppose $h>0$.
\begin{enumerate}[label=(\roman*)]
\item \textbf{No equilibrium leans.} Leaning is strictly dominated, $u_C<0=u_0$, so $x_C=0$ in every
Nash equilibrium. Without an external price anchor, negative feedback is never a rational response to
herd flow.
\item \textbf{Strategic complements.} $u_M(x)$ is strictly increasing in $x$. Hence $x=0$ is an
equilibrium iff $\lambda hG(\lambda h/s)\le\kappa$; $x=\bar X$ is an equilibrium iff
$\lambda h\,q(h,\bar X)\ge\kappa$; and at most one interior equilibrium $x^*$ exists, solving
$u_M(x^*)=0$, and it is unstable. In particular, for
$\kappa\in\big(\lambda hG(\tfrac{\lambda h}{s}),\,\lambda h\,q(h,\bar X)\big)$ both $0$ and $\bar X$
are equilibria.
\item \textbf{Learning tips.} Consider any sign-preserving learning dynamics, with $\dot x_M\gtrless0$
when $u_M\gtrless0$ and $\dot x_C<0$ when $x_C>0$, at speed bounded away from zero wherever the payoff
is. This class includes replicator dynamics and expected softmax policy gradient~\cite{sandholm2010}.
Then $x_C\to0$. If $\lambda h\,q\big(h,x_M(0)-x_C(0)\big)>\kappa$, then $x_M\to\bar X$; if
$\lambda h\,q\big(h,x_M(0)\big)<\kappa$, then $x_M\to0$. With $x_C(0)=0$, the basin boundary is $x^*$.
\item \textbf{Joining amplifies stress.} At threshold $c=0$, down-runs have expected length
\[
\E[L](x)=1+\frac{G\big(\lambda(h+2x)/s\big)}{G(-\lambda h/s)},
\]
which is strictly increasing in $x$. The high-participation equilibrium therefore has strictly more
persistent stress than herding alone.
\end{enumerate}
\end{theorem}

\begin{proof}
(i) From Lemma~\ref{lem:payoff}, $u_C=-\lambda hq-\kappa<0$ because $q>0$ and $\kappa>0$.

(ii) $b$ is strictly increasing in $x$ because $G$ is, and $\partial q/\partial b=(1-a)/(1-a+b)^2>0$,
so $u_M$ is strictly increasing. At $x=0$, $q=a$. The three cases are the equilibrium conditions for an
increasing payoff against the outside option $0$. An interior root is unstable because $u_M$ is
negative below it and positive above it.

(iii) Since $u_C<0$ everywhere, $x_C$ decreases to $0$. Write $w(t)=x_M(t)-x_C(t)$. In the first case
$u_M(w(0))>0$, so $x_M$ rises while $x_C$ falls; then $w$ rises, and since $u_M$ is increasing it stays
positive, so $x_M\to\bar X$. In the second case $u_M(w(0))\le u_M(x_M(0))<0$, so $x_M$ falls; then
$w(t)\le x_M(t)\le x_M(0)$ for all $t$, $u_M$ stays negative, and $x_M\to0$.

(iv) At $c=0$, a down-run starts with a flip, so its first continuation is a fresh period, which
persists with probability $b$; every later continuation is a continuing period, which persists with
probability $a$. Hence $\E[L]=1+b/(1-a)$ with $1-a=G(-\lambda h/s)$, and $b$ is strictly increasing in
$x$.
\end{proof}

\begin{corollary}[What a uniform fee can and cannot do]\label{cor:fee}
A uniform round-trip fee $\kappa>\lambda h\,q(h,\bar X)$ removes every equilibrium with rational
participation. No uniform fee induces leaning: by part (i), $u_C<0$ for every $\kappa\ge0$. Negative
feedback against herds must come from outside this game, for example from arbitrage to an external
reference price or from agents who are paid to take the other side.
\end{corollary}

\begin{proof}
For $\kappa>\lambda hq(h,\bar X)\ge\lambda hq(h,x)$, $u_M(x)<0$ for every $x$, so $x_M=0$ in
equilibrium. The second claim restates part (i).
\end{proof}

Theorem~\ref{thm:together} connects the archetypes. The return-maximizing agent, placed in a pool with
herds, learns to trade ahead of herd flow, which is the behavior the manipulator archetype was designed
to proxy. How a profit-seeking agent behaves depends on who else is in the pool, so the archetypes
cannot be studied in isolation. With $\lambda=s=1$, $h=0.3$, $\kappa=0.2$, and $\bar X=1$, the basin
boundary is $x^*\approx0.21$; softmax policy-gradient learners starting from $15\%$ riders converge to
$0.9\%$, and from $40\%$ to $98\%$, while down-runs lengthen from $2.62$ periods (herds alone) to $3.59$
(Appendix~\ref{app:numerics}).

\section{Adding profit-maximizing market makers}\label{sec:mm}

Corollary~\ref{cor:fee} states that negative feedback must come from agents who are paid to take the other side. We now add them: competitive market makers (MMs) who supply liquidity next to the pool and earn a 
spread for absorbing flow, as wholesalers internalize retail orders. An MM differs from a leaning trader
in one respect only: it is paid to trade rather than paying to trade.

\paragraph{Market makers.} A competitive mass of MMs absorbs a share $\sigma\in[0,\bar\sigma]$ of herd
flow in batch $t$, at the batch midpoint, and earns a net spread $\phi>0$ per unit absorbed. MMs carry
the inventory $-\sigma hD_{t-1}$ for one period and unwind it in batch $t+1$. Riders are as in
Section~\ref{sec:together}, with mass $w=x_M$ (leaning remains dominated for them) and fee $\kappa$.

\paragraph{Prices.} The herd's net impact in batch $t$ is $(1-\sigma)hD_{t-1}$, plus the MMs' unwinding
$\sigma hD_{t-2}$. Adding the riders' round trips,
\[
I_t=\lambda hD_{t-1}+\lambda z\,(D_{t-1}-D_{t-2}),\qquad z=w-\sigma h .
\]
So MMs enter exactly as negative riders, and everything depends on the \emph{net ignition kick} $z$:
rider capital minus the herd flow absorbed by MMs. The fresh-period stay probability becomes
$b(z)=G\big(\lambda(h+2z)/s\big)$, the continuing-period probability $a=G(\lambda h/s)$ is unchanged,
and $q(z)=b(z)/(1-a+b(z))$.

\begin{theorem}[Market makers]\label{thm:mm}
In the stationary regime:
\begin{enumerate}[label=(\roman*)]
\item \textbf{Payoffs.} Riders earn $u_M=\lambda h\,q(z)-\kappa$ per unit; MMs earn
$u_{MM}=\phi-\lambda h\,q(z)$ per unit absorbed. Own impact cancels for both groups.
\item \textbf{Persistence depends only on $z$.} At $c=0$, $\E[L]=1+b(z)/G(-\lambda h/s)$, strictly
increasing in $z$. MMs shorten stress, and when $\sigma h>w$ stress is shorter than with herds alone.
MMs dampen the \emph{start} of a trend but not an established one, since $a$ is unaffected.
\item \textbf{Complements within groups, substitutes across groups.} $u_M$ is increasing in $w$ and
decreasing in $\sigma$; $u_{MM}$ is increasing in $\sigma$ and decreasing in $w$.
\item \textbf{Corner equilibria.} $(\bar X,\bar\sigma)$ is an equilibrium iff
$\kappa\le\lambda h\,q(\bar X-\bar\sigma h)\le\phi$; $(\bar X,0)$ iff
$\lambda h\,q(\bar X)\ge\max(\kappa,\phi)$; $(0,\bar\sigma)$ iff
$\lambda h\,q(-\bar\sigma h)\le\min(\kappa,\phi)$; and $(0,0)$ iff
$\phi\le\lambda hG(\lambda h/s)\le\kappa$. An equilibrium with interior $w$ requires
$\lambda hq(z)=\kappa$, and one with interior $\sigma$ requires $\lambda hq(z)=\phi$; both are unstable
within their own group under sign-preserving dynamics.
\item \textbf{Liquidity collapse and hysteresis.} Let $\kappa<\phi$ and $\lambda hq(-\bar\sigma h)>\kappa$,
so riders are always active, and define $\bar X_c$ and $\bar X_{lo}$ by
\[
\lambda h\,q(\bar X_c-\bar\sigma h)=\phi,\qquad\lambda h\,q(\bar X_{lo})=\phi,
\]
assuming both exist. Since $q$ is strictly increasing, $\bar X_{lo}=\bar X_c-\bar\sigma h$: the
hysteresis band is exactly as wide as the MMs' absorption capacity. For $\bar X<\bar X_c$ the
full-liquidity equilibrium $(\bar X,\bar\sigma)$ exists and is stable. At $\bar X_c$ it disappears, MMs
exit, and $\E[L]$ jumps upward by
\[
\Delta_L=\frac{G\big(\lambda(h+2\bar X_c)/s\big)-G\big(\lambda(h+2\bar X_c-2\bar\sigma h)/s\big)}{G(-\lambda h/s)}>0 .
\]
For $\bar X\in(\bar X_{lo},\bar X_c)$ both $(\bar X,\bar\sigma)$ and $(\bar X,0)$ are stable, so once
liquidity is lost it returns only when rider capital falls below $\bar X_{lo}$.
\end{enumerate}
\end{theorem}

\begin{proof}
(i) Riders follow the proof of Lemma~\ref{lem:payoff} with $w$ replaced by $z$; the $z$ terms cancel.
An MM holds $-\sigma hD_{t-1}$ from midpoint $t$ to midpoint $t+1$, and the same computation gives
gross payoff $-\lambda hq$ per unit, to which the spread $\phi$ is added.

(ii) This is the proof of Theorem~\ref{thm:together}(iv) with $b=b(z)$.

(iii) $q$ is strictly increasing in $z=w-\sigma h$.

(iv) Each condition says that every active group earns at least $0$ and every inactive group at most
$0$. At an interior point, a group's payoff is increasing in its own share, so a small deviation grows.

(v) For $\bar X<\bar X_c$, $\lambda hq(\bar X-\bar\sigma h)<\phi$, and by assumption it exceeds
$\kappa$, so both groups earn strictly positive payoffs at the corner, which is therefore stable. For
$\bar X>\bar X_c$, $u_{MM}(\bar X,\sigma)\le u_{MM}(\bar X,\bar\sigma)<0$ for every $\sigma$, since
$u_{MM}$ is increasing in $\sigma$; so MMs exit, and $z$ jumps from $\bar X_c-\bar\sigma h$ to
$\bar X_c$, and $\Delta_L$ is the difference in (ii). For $\bar X>\bar X_{lo}$,
$u_{MM}(\bar X,0)=\phi-\lambda hq(\bar X)<0$, so $(\bar X,0)$ is a strict, stable equilibrium. On
$(\bar X_{lo},\bar X_c)$ both corners are strict.
\end{proof}

\begin{figure}[t]
\centering
\includegraphics[width=\textwidth]{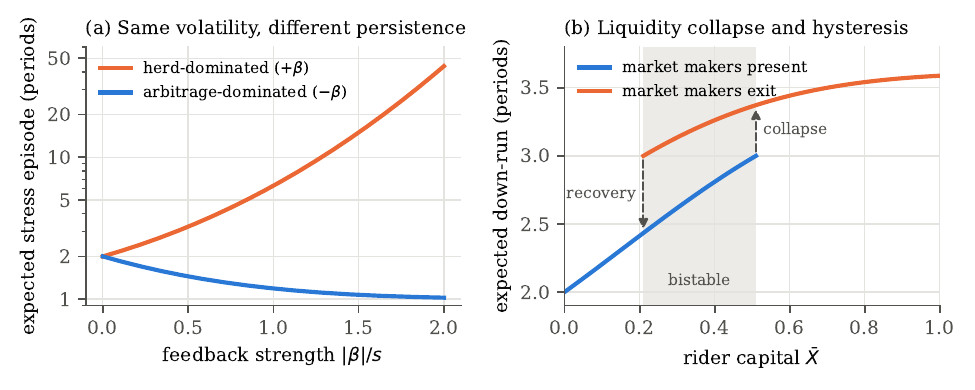}
\caption{\textbf{(a)} Theorem~\ref{thm:main}: expected stress-episode length ($c=0$, Gaussian news) in a
herd-dominated and an arbitrage-dominated market. The two have identical volatility and stress
frequency at every $|\beta|$. \textbf{(b)} Theorem~\ref{thm:mm}(v): expected down-run length against
rider capital $\bar X$ ($\lambda=s=1$, $h=0.3$, $\phi=0.2$, $\bar\sigma=1$). Market makers exit at
$\bar X_c=0.51$ and return only below $\bar X_{lo}=0.21$; in the shaded band both outcomes are stable.}
\label{fig:theory}
\end{figure}

\begin{corollary}[Maker--taker pricing]\label{cor:makertaker}
Corollary~\ref{cor:fee} said no uniform fee induces leaning. With market makers, a direction-blind
\emph{maker rebate} that raises $\phi$ does. Raising $\phi$ raises $\bar X_c$ and $\bar X_{lo}$, which
widens the range of rider capital over which full liquidity is an equilibrium and can be recovered
after a collapse. A taker fee $\kappa>\lambda hq(\bar X-\bar\sigma h)$ removes riders, which sets
$z=-\bar\sigma h$ and minimizes $\E[L]$.
\end{corollary}

\begin{proof}
$q$ is strictly increasing, so $\bar X_c$ and $\bar X_{lo}$ solve $q(\cdot)=\phi/(\lambda h)$ at points
that increase with $\phi$. The taker-fee claim follows from (iv) and (ii).
\end{proof}

\paragraph{What changes with market makers.} Stabilizing behavior now arises in equilibrium: leaning is
still dominated for traders who pay to trade, but profitable for MMs whenever $\phi>\lambda hq$, so what
separates a stabilizer from a non-participant is who pays the spread. Persistence is set by one index,
$z=w-\sigma h$, so each unit of absorption offsets one unit of rider capital, and MMs alone ($z=-0.3$ in
the example of Figure~\ref{fig:theory}b) cut expected down-runs from $2.62$ to $2.00$ periods. The
market becomes fragile: tipping now happens in \emph{liquidity}, with a discontinuous jump in stress
persistence and hysteresis, as in liquidity spirals~\cite{brunnermeier2009} and the Flash
Crash~\cite{kirilenko2017}. And the effective instrument is a maker--taker schedule, which is
behavior-specific, rather than a uniform fee.


\section{Connection to empirical evidence}\label{sec:empirical}

The companion study~\cite{aldridge2026agentic} trains the three archetypes of
Section~\ref{sec:archetypes} with DQN and REINFORCE on hourly bars for AAVE, ETC, and SUI from March
2024 to February 2026, as price-takers in separate environments, and attributes the learned policies
with SHAP. Each policy relies on the signals its objective implies: momentum for the herd-follower,
drawdown and cross-asset returns for the rational agent, and short-term volatility and position for the
manipulator. Three of its findings bear on the theory.

\paragraph{Herders: brief own stress, longer market stress.} The study finds that herd-follower
strategies have the most frequent but shortest stress episodes. This is
Proposition~\ref{prop:herd-own}: with exposure steps of $\tfrac12$, a herd-follower's own episodes last
at most two periods on every path. Theorem~\ref{thm:main}(ii) shows that the same agents lengthen
\emph{market} stress, an effect that strategy-level metrics cannot detect.

\paragraph{Rational agents: stabilizing alone, not together.} The study's rational agents, trading
alone, show the most stable risk-adjusted returns, and its manipulators the rarest but most persistent
stress. Theorem~\ref{thm:together} shows that a return-maximizing agent sharing a pool with herds never
leans against them and, above a computable participation threshold, learns to ride them, which is the
front-running behavior the manipulator archetype proxies, and which lengthens stress. Reading rational
agents as stabilizers is therefore valid only where something else supplies negative feedback, such as
market makers (Theorem~\ref{thm:mm}) or arbitrage to an external price.

\paragraph{What market-level data show.} In the study's data, the two-hour variance ratio is $1.010$--$1.017$, statistically indistinguishable from one. Therefore, by  Proposition~\ref{prop:serial}, the
full sample shows no net feedback at the hourly horizon. Stress nonetheless clusters: runs of hours with returns below $-2\%$ average $1.09$--$1.13$ hours, against $1.02$--$1.04$ under independence, and
absolute returns have autocorrelation $0.22$--$0.27$. This persistence comes from volatility clustering, which Assumption~\ref{ass:noise} excludes. Empirical tests should therefore apply the
serial statistics of Section~\ref{sec:blind} conditionally, relating variation in sign persistence to archetype intensity after controlling for volatility rather than to full-sample levels. The theory
yields three such predictions: herd intensity raises market stress persistence without raising stress frequency at fixed gross flow (Theorem~\ref{thm:main}); the sign of $\mathrm{VR}(2)-1$ tracks the net
index $z$ (Proposition~\ref{prop:serial}, Theorem~\ref{thm:mm}); and market-maker participation falls discontinuously as trend-chasing capital rises and recovers only after a larger decline
(Theorem~\ref{thm:mm}(v)).

\section{Implications for regulation}\label{sec:open}

Corollary~\ref{cor:impossible} shows that limits on volatility, value-at-risk, loss frequency, or turnover cannot distinguish stabilizing from destabilizing populations of agents. Theorem~\ref{thm:detect} shows that the distinction is cheap to read from serial statistics at a rate no method can beat. Corollaries~\ref{cor:fee} and~\ref{cor:makertaker} show that a uniform fee can deter riders but never induce stabilizing trades, while a maker rebate does and widens the range of trend-chasing capital over which liquidity survives and can be restored. Instruments should act on feedback rather than on one-period statistics.

\section{Conclusion}\label{sec:conclusion}

Defining trading agents by their objectives traces a path from rewards to order flow to market stress.
On that path, the one-period statistics regulators watch are provably blind, while serial statistics
separate stabilizing from destabilizing feedback at an optimal rate. When agents trade together,
profit-seeking agents amplify herds, and market makers that are paid to absorb flow restore stability
at the cost of fragile, hysteretic liquidity. Instruments that act on feedback, such as maker--taker
pricing, are the effective response.

\bibliographystyle{plain}
\bibliography{references}

\clearpage
\appendix
\section{Numerical verification}\label{app:numerics}

Every closed form in Sections~\ref{sec:blind}--\ref{sec:mm} was checked by Monte Carlo simulation with
Gaussian news and $\lambda=s=1$. Table~\ref{tab:numerics} reports simulated values with theoretical
values in parentheses.

\begin{table}[h]
\centering
\small
\begin{tabularx}{\textwidth}{@{}lX@{}}
\toprule
Result and setting & Simulated (theory)\\
\midrule
Thm.~\ref{thm:main}, $\beta=+0.8$, $c=1$, $2\times10^6$ periods & $\Var(r)$ $1.6398$ ($1.6400$); $F$
$0.2282$ ($0.2283$); $\E[L]$ $1.7247$ ($1.7263$); $q$ $0.7883$ ($0.7881$)\\
Thm.~\ref{thm:main}, $\beta=-0.8$, $c=1$ & $\Var(r)$ $1.6398$ ($1.6400$); $F$ $0.2287$ ($0.2283$);
$\E[L]$ $1.0372$ ($1.0373$); $q$ $0.2117$ ($0.2119$)\\
Prop.~\ref{prop:herd-own}, $\Delta=\tfrac12$, $3\times10^5$ periods & longest own-stress episode $2$
($\le2$); mean $1.08$\\
\midrule
Lemma~\ref{lem:payoff}, $h=0.3$, $(x_M,x_C)=(0,0)$, $10^6$ periods & rider gross $0.1847$ ($0.1854$);
$q$ $0.6181$ ($0.6179$); $\E[L]$ $2.6164$ ($2.6172$)\\
\quad $(x_M,x_C)=(0.5,0)$ & $0.2102$ ($0.2108$); $0.7026$ ($0.7027$); $3.3608$ ($3.3638$)\\
\quad $(x_M,x_C)=(1,0)$ & $0.2155$ ($0.2164$); $0.7215$ ($0.7214$); $3.5860$ ($3.5891$)\\
\quad $(x_M,x_C)=(0,0.5)$ & $0.1173$ ($0.1163$); $0.3874$ ($0.3877$); $1.6295$ ($1.6333$)\\
Thm.~\ref{thm:together}(iii), $\kappa=0.2$, $\bar X=1$, softmax policy gradient & from shares
$(0.15,0.15,0.70)$: $(0.009,0.001,0.991)$; from $(0.40,0.15,0.45)$: $(0.981,0.001,0.018)$; basin
boundary $x^*\approx0.21$\\
\midrule
Thm.~\ref{thm:mm}(i)--(ii), $(w,\sigma)=(0,0.5)$, $z=-0.15$ & rider gross $0.1699$ ($0.1701$); MM gross
$-0.1699$ ($-0.1701$); $\E[L]$ $2.308$ ($2.309$)\\
\quad $(w,\sigma)=(0.5,1)$, $z=0.20$ & $0.1988$ ($0.1995$); $-0.1988$ ($-0.1995$); $2.996$ ($2.984$)\\
\quad $(w,\sigma)=(1,0.5)$, $z=0.85$ & $0.2168$ ($0.2157$); $-0.2168$ ($-0.2157$); $3.571$ ($3.558$)\\
Thm.~\ref{thm:mm}(v), $\kappa=0.12$, $\phi=0.2$, $\bar\sigma=1$ & $\bar X_c=0.510$, $\bar X_{lo}=0.210$;
$\E[L]$ jumps $3.000\to3.373$; learning dynamics lose MMs at $\bar X\approx0.6$ on the way up and
recover them at $\bar X\approx0.1$ on the way down\\
\bottomrule
\end{tabularx}
\caption{Simulated versus theoretical values. Payoffs are gross of fees, per unit per period.}
\label{tab:numerics}
\end{table}

\paragraph{Simulation design.} Section~\ref{sec:blind} simulates~\eqref{eq:model} directly. Sections
\ref{sec:together}--\ref{sec:mm} simulate the batch model with impact $I_t$, fill each order at the
batch midpoint, and compute payoffs on round trips from batch $t$ to $t+1$. For
Theorem~\ref{thm:together}(iii), each of 400 iterations estimates the three payoffs from a fresh
simulation of $2\times10^4$ periods at the current shares and applies the expected softmax
policy-gradient update with step size $4$. For Theorem~\ref{thm:mm}(v), rider and MM shares follow
logistic sign-preserving dynamics on the closed-form payoffs, run to convergence at each $\bar X$ while
$\bar X$ is swept up from $0.05$ to $0.8$ and back. Learning slows near each threshold, so the
simulated collapse and recovery points lie outside the band $[\bar X_{lo},\bar X_c]$ rather than on its
edges, as stability predicts. The code is provided as supplementary material.

\end{document}